# Slow electron attachment as a probe of cluster evaporation processes

Roman Rabinovitch,[1] Klavs Hansen,[2] and Vitaly V. Kresin[1]

[1] *Department of Physics and Astronomy, University of Southern California,*

*Los Angeles, California 90089-0484, USA*

[2] *Department of Physics, University of Gothenburg, SE-412 96 Gothenburg, Sweden*

**Abstract**

Neutral alkali clusters efficiently capture low-energy electrons with the aid of long-range polarization attraction. Upon attachment, the electron affinity and kinetic energy are dissipated into vibrations, heating the cluster and triggering evaporation of atoms and dimers. This process offers a novel means to explore nanocluster bonding and evaporation kinetics. The present work investigates the formation of $Na_N^-$. A crossed-beam experiment reveals that relative anion abundances become strongly and nontrivially restructured with respect to the neutral precursor beam. This restructuring is explained in quantitative detail by an analysis of evaporative cascades initiated by the attachment. The analysis thus furnishes a complete description of the electron attachment process, from initial attraction to final rearrangement of the cluster population. In addition, the paper describes a systematic derivation of cluster evaporation kinetics and internal temperature distributions; a new relation between the dissociation energies of cationic, neutral and anionic metal clusters; and a scenario for inferring the neutral cluster population in the supersonic beam from the cationic mass spectrum.

## I. Introduction

Fundamental insights into structures and stabilities of cluster nanoparticles have been derived from their abundance spectra, i.e., from the intensity patterns in the mass spectra of size-resolved cluster beams.[1-5] For example, clusters of alkali metals, which are a benchmark for studying the electronic structure of metal nanoparticles, famously display "magic numbers" in their mass spectra:[4,6,7] prominent peaks or steps in intensities corresponding to sizes 8,20,40,58,92,138,... When first discovered,[8] this sequence was immediately interpreted in terms of electronic shell structure, i.e., the enhanced stability of particles whose number of delocalized valence electrons corresponds to the completion of an energy shell in the self-consistent potential well. What was not immediately clear at the time is how these spectra are generated in the experimental device and the precise mechanism that causes the mass spectra to reflect the shell closings.

For hot-nozzle forces, the abundances are now understood to derive not from a quasi-equilibrium distribution but from post-production evaporative processes: vapor expansion and cluster nucleation are followed by collision-free flight during which the initially hot clusters decay in vacuum. For alkali clusters the dominant channels for these decays are evaporation of atoms and dimers.[9,10] The evaporation rate constants depend very strongly on the internal energy of the clusters and on their dissociation energy. Therefore the measured intensity patterns reflect the result of a sequence of decay steps, and the magic-number maxima originate from evaporative bottlenecks encountered at particularly stable cluster sizes.

Although the nanocluster energy and temperature distributions,[11] size-to-size abundance variations, lifetimes, etc. do not accrue from standard equilibrium thermodynamics, they can be analyzed in detail using the so-called "evaporative ensemble" formalism, a term introduced by Klots.[12]  Starting from an ensemble of initially highly excited clusters and assuming statistical equipartition, it reveals that the distributions realized downstream are highly constrained by the observation time and by a few inherent properties of the particles, primarily their relative dissociation energies.  The formalism possesses a gratifying universality, and has been successfully applied to systems ranging from $T \approx 0.4$ K liquid helium nanodroplets[13], to alkali metal clusters at $T \approx 400$ K,[14,15] to $C_{60}$ beams originating at $T \approx 4000$ K.[16]

On the other hand, this universality prevents one from straightforwardly extracting some of the most important cluster parameters directly from the abundance spectra:  for example, because the relative abundances are determined by the ratios of the dissociation energies,[17] the absolute magnitudes of the latter cannot be determined without the presence of an absolute energy scale in the problem.  Such a scale can be introduced, for example, by reheating the particles downstream from the nozzle with a well-defined excitation energy.  This shifts their internal energy distribution upwards and causes a second burst of evaporation, resulting in a restructuring of the abundance spectrum.  Connecting this restructuring with an understanding of the excitation process refines one's knowledge of the clusters' properties.  In alkali-cluster work, such re-excitation experiments have utilized ionization with an intense laser pulse,[9,10,18] photoabsorption,[15,19,20] charge transfer,[21,22] and sticking collisions of atoms with clusters.[23]

Here we present the application of another probe:  attachment of low-energy electrons to neutral sodium clusters.  Previous experiments with alkali-metal clusters have shown that free electron capture leads to efficient formation of anions.[24,25]  This is due to the high polarizability of the clusters[4,26-28] which gives rise to strong long-range forces in interaction with charged particles[29] and to very high electron capture cross sections for $E_e \ll 1$ eV.[30,31]

The formation of stable negative species is accompanied by the release of electron kinetic and affinity energy. If the resultant rearrangement of the abundance spectrum is weak, the daughter anions will have maximal intensities at the peaks of the neutral precursor spectrum (i.e., 20,40,58,… atoms). Alternatively, the post-attachment spectra may shift to shell closings at $Na^-_{19,39,57,\ldots}$. As will be shown below, the major magic numbers do indeed shift in the experiment, but the daughter anion mass spectrum is by no means a simple uniform displacement of the precursor pattern by one atom. In fact, interpretation of the observed anion distribution requires a thorough treatment of the evaporative ensemble dynamics. In return, one attains a complete description of the electron attachment process – from initial attraction to final rearrangement –as well as improved information about the dissociation energies of neutral and ionic cluster species. A brief report of this work was given in a previous publication.[32] Here we present a full account of the analysis and its results, in a comprehensive illustration of the use of evaporative ensemble theory to treat the evolution of cluster populations.

Sec. II outlines the electron attachment experiment, and Sec. III displays the resulting anion abundance spectra and highlights those features which call for a detailed analysis of the accompanying evaporation kinetics. The roadmap of this analysis is presented in Sec. IV, and its implementation in Sec. V. As demonstrated at the end of Sec. V and discussed in Sec. VI, the calculation yields a very good explanation of the experimentally observed anion abundance pattern, validating the evaporative attachment scenario.

Some important material has been relegated to the appendices. Appendix A addresses the issue of neutral cluster population in the supersonic beam by reconstructing a fragmentation-free precursor mass spectrum. Appendices B and C assemble a systematic derivation of the atom and dimer evaporation rates of hot clusters, and of the clusters' internal temperature distributions.

Appendix D formulates a new relation between the dissociation energies of neutral and ion clusters.

## II. Experimental procedure

An outline of the experimental arrangement is shown in Fig. 1; a detailed description can be found in Refs. 32,33. A beam of neutral sodium clusters generated in a supersonic expansion source is intersected at a right angle by a ribbon of low-energy electrons (~10 μA current, average kinetic energy $\bar{E}_{e^-}$ =0.1 eV[34]). Due to the low densities of both beams, the probability of multiple collisions is negligible. An approaching electron polarizes a cluster and can become captured by the resulting dipole moment. The corresponding "Langevin" [24,35] cross section is given by

$$\sigma_{capture} = \sqrt{\frac{2\pi^2 e^2 \alpha}{E_{e^-}}},$$  (1)

where $\alpha$ is the cluster's electric polarizability.[36] For Na$_{20}$, Na$_{40}$, and Na$_{58}$ this corresponds to $\sigma_{capture}$≈1400 Å$^2$, 2100 Å$^2$, and 2500 Å$^2$, respectively. The clusters are efficient at dissipating the attachment energy, as analyzed below, hence an electron that, classically, reaches the cluster surface can be assumed to have 100% sticking probability.

Downstream from the collision region, ion optics guide the formed Na$_N^-$ products into a quadrupole mass spectrometer (Extrel QPS-9000). Its channeltron detector was specially engineered (DeTech Inc.) for the effective detection of heavy negative ions which requires a significantly higher voltage applied to the conversion dynode[38] in our case 16 kV. The use of low noise power supplies and careful preconditioning of the dynode allowed us to reach a detector dark count noise below 1 count per second (cps). This is essential, because despite the use of an electron gun designed for high currents at low energies[39] and optimization of the supersonic source flux and ion optics settings, only a small fraction of the cluster beam undergoes a charging collision. In combination with the deflecting action of the electron gun's internal magnetic field and a ~3% channeltron detection efficiency,[24] this resulted in typical anion signal intensities for an individual cluster mass ranging from only a few up to ~$10^2$ cps.

The part of the cluster beam that remains neutral proceeds undisturbed into the rear chamber of the setup where the particles are ionized by focused ultraviolet light, mass selected by a second quadrupole mass spectrometer, and detected with an ion counter. Various measurements of size-dependent properties of alkali metal clusters have indicated that mass spectra produced by filtered near-threshold UV ionization closely reflect the population of the original neutral beam;[4] a further refinement is presented in Appendix A. The ability to record both the anion (daughter) and the neutral (precursor) cluster abundances simultaneously is a key aspect of the present experiment, because it makes it possible to follow and analyze the relative changes in cluster populations without distortions due to temporal beam variations.

### III. Measured abundance distributions

The anion mass spectra were acquired in segments, with experimental conditions optimized for maximum signal within the chosen mass range.[33] Combined, the data covered the cluster size range from $Na_7$ to $Na_{92}$ and from $Na_{132}$ to $Na_{144}$, as shown in Fig. 2. It reveals a significant amount of restructuring with respect to the neutral parent population (as mirrored by the measured cation spectrum, see the preceding paragraph):

(i) The overall envelope of the anion spectrum is moved towards the higher masses.

(ii) The abundance magic numbers are shifted from $N$=(8, 20, 40, 58, 92, and 138) in neutral Na$_N$ clusters to $N$=(7, 19, 57, 91, and 137) in Na$_N^-$, reflecting the electron shell closings in anions. Correspondingly, the "closed-shell-plus-one-electron" clusters display a low intensity, in particular the signal of Na$_8^-$ is too low to be observed and the peak of Na$_{20}^-$ is very small.

(iii) On the other hand, the "closed-shell-minus-one-electron" anions have higher relative abundances than the corresponding neutrals.

(iv) In between the magic numbers, the relative intensities of the anion peaks are in an inverse correlation with the intensities of the peaks of the neutrals, in a way which is not a simple shift by one electron number.

While the envelope change could be rationalized as due to different detection efficiencies of the anions and of the precursor cluster beam, and the magic number shift is intuitively attributable to the extra acquired electron, understanding the shifts and other changes in the abundance spectra quantitatively requires a more detailed analysis which will be presented below.

## IV. Outline of the evaporative attachment scheme

The detection of anions implies that the capture of electrons by the cluster polarization potential is followed by the formation of a bound state between the cluster and the electron. In principle, the electron could be captured in the potential formed by the polarization and centrifugal potentials, but it is unlikely that such a metastable state would survive for the ~$10^{-4}$ s duration required for the experimental detection.[40] The only other possibility is that the electron falls into an available shell state inside the cluster. The electron affinity can be dissipated either radiatively, or as vibronic and ultimately vibrational excitation energy. Theoretical studies have shown that the radiative channel is weak for collision energies below the collective electron resonances, i.e., below several eV.[42]   This means that the primary outcome of the electron capture process is a strong rise in the randomized vibrational excitation of the cluster, i.e., heating.

Even before the collision, the clusters are hot enough to be molten.[43-45]   Using the formalism and parameters described below, one finds that the dissipation of an additional 1.2 eV (a lower limit for the Na cluster electron affinity[46-48]) further increases the internal temperatures of $Na_8$, $Na_{20}$, and $Na_{40}$ by ≈580 K, ≈230 K, and ≈120 K, respectively, thereby causing their unimolecular decay rate to increase by approximately 5, 3, and 2 orders of magnitude.   As a result, a certain amount of fragment evaporation within the post-attachment flight time will be unavoidable.   Such statistical "electron capture dissociation"[49] processes have been labeled[50] "evaporative attachment."   The result is an evaporative cascade: the anions cool by evaporating Na atoms ("monomers") and $Na_2$ dimers.   In addition, the dissociation pathways of the hot anions are strongly affected by the fact that the additional valence electron alters, in a nonmonotonic way, the dissociation energy of a cluster and thereby its decay rate constant and branching pattern.   These effects are at the root of the nontrivial restructuring of the abundance spectrum.   The cascade process is illustrated schematically in Fig. 3.

The quantitative treatment comprises the following steps.

(I) First, the vibrational temperature (energy) distributions of the neutral precursor clusters are deduced by employing the evaporative ensemble approach, based on the clusters' dissociation energies and beam flight times.

(II) Second, the temperature distributions of the cluster anions immediately following the attachment are calculated by using the vibrational energy distributions of the precursor clusters, the kinetic energy of the electrons, and the cluster electron affinity values.

(III) Third, the evaporative ensemble formalism is used to evaluate the evaporation pathways of the parent hot anions and the relative abundances of the daughters.

(IV) Finally, the calculated fragmentation patterns are convoluted with the mass spectra of the neutral precursor clusters (measured in parallel), with the Langevin electron attachment cross sections, and with the geometrical ion transmission function of the electron gun.

These steps are detailed in the next section.

## V. Analysis

### A. Evaporation kinetics

Thermal evaporation rate constants of free nanoclusters can be derived based on the theory of detailed balance, a formalism originally developed for nuclear reactions by Weisskopf[51,52] and later adapted to cluster systems (see, e.g., Refs. 12,17,53-57). Advantages of this approach are that monomer and dimer evaporation are treated similarly on a rigorous basis, and no RRKM-type[58] assumptions about transition states are required.

The details of the formulation are reviewed in Appendix B. The main result [Eqs. (B.14), (B.17)] is that the rate constant for the evaporation of an atom, denoted "$m$" for "monomer" {or a dimer, denoted "$d$"} by a parent cluster of $N$ atoms and temperature $T$, can be written in the approximate Arrhenius form:

$$k_N^{m\{d\}}(T) \approx \omega^{m\{d\}} N^{2/3} \exp\left(-D_N^{m\{d\}} / k_B \tilde{T}_N^{m\{d\}}\right) \qquad (2)$$

Here $D^{m\{d\}}$ is the dissociation energy, i.e., the minimum energy required to remove an atom {or an intact dimer} from the cluster surface. (The term "dimer dissociation energy" will be used in this context throughout and should not be confused with the energy required to break apart the dimer molecule.) The effective temperatures $\tilde{T}$ and the prefactors $\omega$ are defined in Appendix B.

The probability to evaporate a fragment within a defined time interval can be easily obtained from the $k(T)$. Considering an ensemble initially consisting of $n_N(0)$ clusters of size $N$ at temperature $T$, the number of clusters undergoing monomer or dimer evaporation within a time interval $dt$ satisfies $dn = -n\left[k_N^m(T) + k_N^d(T)\right]dt$, from which it follows that the number of parents remaining intact after a time $t$ is given by $n_N(t) = n_N(0)\exp\left\{-\left[k_N^m(T) + k_N^d(T)\right]t\right\}$ and, correspondingly, the probability of a cluster having lost a monomer (dimer) by that point is

$$P_N^{m\{d\}}(T,t) = \frac{k_N^{m\{d\}}(T)}{k_N^m(T) + k_N^d(T)}\left(1 - e^{-\left[k_N^m(T) + k_N^d(T)\right]t}\right). \qquad (3)$$

To illustrate the character of these functions, an example is shown in Fig. 4. Note that due to the difference in pre-factor values [see Eq. (B.18)], the evaporation rate constants for the two branches can be of the same order of magnitude at the same temperature even when the dimer dissociation energy is higher than that of monomer.

With the distribution of internal temperatures of clusters of size $N$ denoted by $F_N(T)$, the total fraction that undergoes either evaporation scenario within a time $t$ is given by

$$f_N^{m\{d\}}(t) = \int P_N^{m\{d\}}(T,t) F_N(T) dT \qquad (4)$$

*B. Temperature distributions*

To determine the fragmentation pathways of hot cluster anions, we therefore need to know their internal temperature distribution [see Eq. (4)], and this in turn requires a knowledge of the original temperature distribution of the neutral precursors. This matter is considered in Appendix C, and the result is that the precursors' distribution $F_N^{(neutral)}$ is taken to be either (*i*) a flat rectangle bounded by $T_{\min}^{(neutral)}$ and $T_{\max}^{(neutral)}$, given by Eqs. (C.2) and (C.3) and shown in Fig. 9(a), for cases when the precursor itself originated via monomer evaporation; or (*ii*) a double rectangle, as discussed at the end of Appendix C and illustrated in Fig. 9(b), for cases when the precursor could be formed either by monomer or by dimer evaporation channels. The pertinent branching ratios were assumed to be the same as determined for isoelectronic cationic clusters in Ref. 9, and the neutral clusters' dissociation energies for monomer and dimer evaporation were derived from the same reference as discussed in Appendices D,E.

Assuming that the captured electron promptly ends up in the lowest available single-particle electronic state, its full excess energy is transferred to the vibrational heat bath, so that

$$T_N^{(-)} = T_N^{(neutral)} + \frac{\overline{E}_{e^-} + A_N}{(3N-7)k_B}. \tag{5}$$

The numerator is the sum of the electron's incoming kinetic energy (see Sec. II) and the cluster's electron affinity (here approximated by the experimentally measured detachment energies of cluster anions[46-48]), and the denominator is the microcanonical heat capacity, Eq. (B.11). Thus the temperature distributions of the neutral precursors $F_N^{(neutral)}\left(T_N^{(neutral)}\right)$ are shifted upwards by the above amount, giving us the temperature distributions of the anions $F_N^{(-)}\left(T_N^{(-)}\right)$, plotted in Fig. 5. These are likewise flat rectangular distributions, which simplifies the calculations.

These heated anions now evaporate *en route* to the detector, giving rise to the experimentally observed $Na_N^-$ abundance spectra.

*C. Anion fragmentation patterns*

The full evaporation chain in Fig 3(b) is now to be computed for every initial cluster anion size. We start with its temperature distribution at inception, $F_N^{(-)}\left(T_N^{(-)}\right)$, and use Eqs. (3), (4) to enumerate how many daughters it will engender by monomer and dimer evaporation during the flight time from the electron gun to the channeltron detector. In the present experiment, this flight time ranged from 0.3 ms to 1.2 ms depending on the cluster size.[59] The anion dissociation energies used in the calculation are again derived from the data in Ref. 9 as discussed in Appendices D and E.

Some daughters will be sufficiently hot to evaporate yet again. To trace the contribution of these hot daughters we need once more to know their internal temperature distributions. This is taken into account by comparing the threshold evaporation temperature of the most probable channel of the daughter's decay and assuming that the population above this temperature will decay again [thus losing another quantity of energy $\Delta E$, Eq. (C.1)] and the fraction below this temperature will not.

As illustrated in Fig. 3(b), each channel is traced until every product has a negligible probability of decay during the detector flight time, i.e., until the entire width of its temperature distribution lands below the threshold temperature. These end products, circled in the figure, form the fragmentation pattern. The calculations show that the majority of anions in the considered range undergo no more than two consecutive evaporations. The only significant exceptions are $Na_{22}^-$ and $Na_{23}^-$ for which half of the population evaporates three monomers, or a dimer followed by two monomers, respectively.

### D. Results

The initial population of hot parent anions is obtained by convoluting the mass spectrum of the neutral cluster beam (see Fig. 7 in Appendix A), with the size-dependent electron attachment cross section, Eq. (1). Next, evaporative cooling chains are calculated for every anion, as described above. We find that in order to reproduce the observed products, it is important to incorporate both atom and dimer loss pathways. The resulting distribution is then weighed by an instrumental scaling factor $\sqrt{N}$ : the relative probability for a charged cluster to traverse the electron gun collision region and reach the entrance lens of the quadrupole without being deflected by the magnetic field inside the gun.[24] The end result may now be compared with the experimental anion mass spectrum.

The comparison is given in Fig. 6 for clusters up to $Na_{33}^-$ – the largest size for which evaporation cascades can be fully mapped from the dissociation energy data in Ref. 9. The calculation reproduces the experimental pattern very well, including the nontrivial intensity variations of open-shell clusters pointed out in Section III, and the peculiar feature that the $Na_{18}^-$ peak is stronger than the closed-shell $Na_{19}^-$. The latter is due to extensive evaporation of the parent $Na_{20}$.

We conclude that the scheme described in this paper provides an accurate, unified portrait of negative ion formation by low-energy electron attachment. Efficient capture by the long-range polarization force is followed by thermalization of the energy deposited by the incoming electron's energy, and the abundance distribution becomes strongly rearranged by evaporative emission of atom and dimer fragments.

It is interesting to note that in an experiment on electron transfer from $Kr^{**}$ atoms to neutral potassium clusters,[60] the structure of the $K_{3-32}^-$ distribution was not unlike that found here for $Na_N^-$. The authors of Ref. 60 hypothesized that the pattern was due to strong cluster-shape related oscillations in the attachment cross section, and did not consider post-transfer evaporation effects. It follows from our work that in actuality the cross sections do not display dramatic size-to-size shifts, and the abundance variations are caused by "evaporative electron transfer."

## VI. Conclusions

An experimental measurement reveals a nontrivial transformation of cluster abundance mass spectra upon electron capture by the strong polarization force. The kinetics of the transformation are explained by a careful evaluation of cluster heating resulting from the incoming electron's energy dissipation, and the ensuing monomer and dimer evaporation cascades. The fact that the temperature rise and the evaporation pathways are strongly size-dependent lies at the root of the restructuring of the ion abundances. The close agreement between the measured and calculated anion abundance spectra, achieved without adjustable parameters, affirms the general scenario of the process and the methodical application of evaporative ensemble theory. It also supports the relation, formulated in Appendix D, between the neutral and ion cluster dissociation energies.

Since evaporative cooling is exponentially sensitive to temperatures and dissociation energies, slow-electron capture offers a useful window into the statistical and binding properties of clusters, as well as molecules with strongly coupled vibrational modes.

Evaporation will remain an important channel on the experimental time scale until nanoclusters grow to be too massive to heat up appreciably. An estimate gives $N \sim 10^3$ as the Na cluster size at which the anion mass spectrum should begin to mirror that of the neutral precursor beam. For attachment of more energetic electrons, evaporation will recur. It would be quite interesting to map out such variations, as well as the competition between radiation and evaporation channels, as a function of electron energy and nanocluster size.

## Acknowledgments

This work was supported by the U.S. National Science Foundation (PHY-0652534) and by the Swedish Science Research Council (VR).

**Appendix A: Mass spectra and abundances of neutral clusters**

How closely does the mass spectrum of an alkali cluster beam ionized by "soft" filtered UV lamp ionization resemble the abundance of neutral precursor clusters? This question is important not only for the modeling of evaporation cascades in the present experiment, but for a multitude of other measurements of size-dependent cluster properties. As mentioned in Sec. II, a large amount of experimental data are consistent with the supposition that there is a close correspondence between the ionized and neutral populations. In this appendix, we give additional consideration to this matter.

We begin by noticing something peculiar about the peaks at $n=9$ and $n=21$ in the mass spectrum shown in Fig. 2(b). According to the branching ratios determined in Ref. 9, there is no evaporation process that could lead to the formation of *neutral* clusters of these sizes. Indeed, clusters with 10 and 22 valence electrons decay only by evaporating dimers, while clusters with 11 and 23 valence electrons evaporate only monomers. Therefore neutral $Na_9$ and $Na_{22}$ would be present in the incoming beam only if they were generated in the nozzle expansion in a state of low internal temperature and arrived at the detector without having undergone any evaporation *en route*. The likelihood of this is low, therefore we infer that the $n=9,21$ peaks in the UV mass spectra in Fig. 2(b) are a consequence of direct cluster fragmentation induced by the lamp. This conclusion is supported by contrasting these relatively strong peaks with the very small or non-existent abundances of isoelectronic cations $Na_{10}^+$ and $Na_{22}^+$ in the mass spectra shown in Refs. 9,10, and of anions $Na_8^-$ and $Na_{20}^-$ in Fig. 2(a).

To take into account the effect of partial direct fragmentation of clusters, we introduce the following correction. Instead of taking the measured mass spectrum $M^{(lamp)}(N)$, produced by the UV lamp, to be completely identical to that of the incoming neutral clusters, we treat it as a superposition of two distributions: that of the authentic unfragmented neutral precursors $M^{(neutral)}(N)$, and that of cations which have been restructured by ionization-induced fragmentation and evaporation, $M^{(restructured)}(N)$. That is, $M^{(neutral)}(N)=M^{(lamp)}(N)-M^{(restructured)}(N)$. The exact shape of the restructured distribution is unknown; however, we shall assume that it resembles the one generated by ionizing the same cluster beam with 350 nm pulses from a tunable ns laser, shown in the insert in Fig. 7. Indeed, such intense above-threshold pulses should result both in significant cluster heating and in fragmentation,[10,61] and the distribution we observed indeed strongly resembles, for example, that in Refs. 9,10. Therefore, calling this distribution $M^{(laser)}(N)$, we make the approximation that $M^{(restructured)}(N)$ parallels it to within a smoothly varying envelope: $M^{(restructured)}(N) \approx (aN+b)M^{(laser)}(N)$. The coefficients $a$ and $b$ are fixed by requiring that the resultant $M^{(neutral)}(9)=M^{(neutral)}(21)=0$, in accordance with the above discussion. Obviously, this is an inexact representation, but it provides a guide to the magnitude of the correction in question.

Adjusting the original mass spectrum in Fig. 2(b) in this manner, we obtain the revised neutral cluster abundance distribution presented in Fig. 7. Gratifyingly, it shows that the modification for most cluster sizes is minor.

# Appendix B:  Overview of evaporation statistics

Appendices B and C assemble a systematic description of those aspects of the Weisskopf formalism for cluster evaporation that are essential for the present analysis.  The formalism is based on the principle of detailed balance: the evaporation rate of a parent cluster is equal to the rate of the reverse process (attachment of a fragment to a daughter), $k^{(evap)}\rho_{parent} = k^{(att)}\rho_{products}$. Here $\rho_{parent}$ is the density of states of the parent cluster, and the differential density of states of the products is the convolution of the internal state density of the daughter cluster with those of the translational and internal (if any) degrees of freedom of the evaporated small fragment:

$$\rho_{products}\,d\varepsilon_t\,d\varepsilon_{\text{int}} = \rho_{daughter}\left(E_{parent} - D^{m\{d\}} - \varepsilon_t - \varepsilon_{int}^{m\{d\}}\right)\cdot\rho_t\left(\varepsilon_t\right)d\varepsilon_t\cdot\rho_{\text{int}}\left(\varepsilon_{int}^{m\{d\}}\right)d\varepsilon_{\text{int}} .$$ 

(B.1)

In this equation $\varepsilon_{int}^{m\{d\}}$ is the total internal energy of the monomer {dimer} fragment, $\varepsilon_t$ is its translational kinetic energy, and $D^{m\{d\}}$ is the dissociation energy.  This yields the following expression for the differential evaporation rate constant:[62]

$$k_N^{m\{d\}}\left(E_N, \varepsilon_{int}^{m\{d\}}, \varepsilon_t\right) = \frac{g^{m\{d\}}\mu^{m\{d\}}\sigma\varepsilon_t}{\pi^2\hbar^3}\cdot\frac{\rho_{N-1\{N-2\}}\left(E_N - D_N^{m\{d\}} - \varepsilon_{int}^{m\{d\}} - \varepsilon_t\right)\rho^{m\{d\}}\left(\varepsilon_{int}^{m\{d\}}\right)}{\rho_N\left(E_N\right)} .$$

(B.2)

As defined above, $E_N$ and $D_N^{m\{d\}}$ are the internal and dissociation energies of the parent cluster; $\sigma$ is the cross section of the reverse (association) process, $g$ is the fragment spin degeneracy, and $\mu$ is the reduced mass of the fragment+daughter system.

In the temperature range of interest, the excited electronic states of neither the clusters nor the fragments are accessible, and the internal state densities are those of the vibrational excitations.  (In a more general situation, one would need to include the electronic partition functions.[63])  Since rotational state densities of the heavy clusters are very high, the discreteness and conservation of angular momentum do not impose any additional limitations (for a more general case see Ref. 64). Indeed, the classical rotational level densities of the parent and the daughter cancel to a good approximation provided their angular momenta are significantly greater than that of the decay channel,[64] as will be assumed here. We will, however, need to account for the rotational state densities of the light dimer fragments (see below).

*Monomer evaporation*. Since atoms do not have rotational and vibrational degrees of freedom, and the electronic excitations are negligible in the present situation, the monomer evaporation rate constant simplifies:

$$k_N^m(E_N, \varepsilon_t) = \frac{2M\sigma\varepsilon_t}{\pi^2\hbar^3} \frac{\rho_{N-1}(E_N - D_N^m - \varepsilon_t)}{\rho_N(E_N)}. \tag{B.3}$$

Here $g^m = 2$ and $\mu^m \approx M$ are the Na atom spin degeneracy and mass, and the atom-cluster sticking cross section will be taken approximately equal to the hard-sphere value $\sigma \approx \pi r_s^2 N^{2/3}$ ($r_s$ is the metal's Wigner-Seitz radius). It is convenient to expand the density of states in the numerator as follows:

$$\rho_{N-1}(E_N - D_N^m - \varepsilon_t) = e^{\ln \rho_{N-1}(E_N - D_N^m - \varepsilon_t)} \approx \rho_{N-1}(E_N - D_N^m)e^{-\varepsilon_t/k_B T_{N-1}} \tag{B.4}$$

We have introduced the "internal temperature of a cluster" as a convenient parameter:

$$\frac{1}{k_B T_{N-1}} = \frac{d \ln \rho_{N-1}(E)}{dE}\bigg|_{E=E_N - D_N^m}, \tag{B.5}$$

and made use of the fact that the average value of the kinetic energy release $\langle \varepsilon_t \rangle \ll E_N - D_N^m$. Indeed, from (B.4)

$$\langle \varepsilon_t \rangle = \left(\int \varepsilon_t e^{-\varepsilon_t/k_B T_{N-1}} d\varepsilon_t\right)\bigg/\left(\int e^{-\varepsilon_t/k_B T_{N-1}} d\varepsilon_t\right) = 2k_B T_{N-1}, \tag{B.6}$$

which in our case ($T \lesssim 1000$ K prior to the onset of evaporation) is much less than the total vibrational energy and the dissociation energy.

Integrating the resulting expression for the differential monomer evaporation rate over $\varepsilon_t$ from 0 to $E_N - D_N^m$ we obtain[51,52]

$$k_N^m (E_N) = \frac{2 M r_s^2 N^{2/3} (k_B T_{N-1})^2}{\pi \hbar^3} \frac{\rho_{N-1}(E_N - D_N^m)}{\rho_N(E_N)} . \qquad (B.7)$$

This can be simplified further by making the "Einstein" approximation that the hot cluster's vibrational density of states is close to that of an ensemble of $s = 3N-6$ harmonic oscillators of frequency $\hbar \omega_0$, which is given by the Kassel equation:[58] $\rho_s(E) = E^{s-1} / \left[ (s-1)! (\hbar \omega_0)^s \right]$. Then in the numerator of Eq. (B.7) we can substitute $\rho_{N-1} \left( E_N - D_N^m \right) \approx \left( \Theta_D / T_{N-1} \right)^3 \rho_N \left( E_N - D_N^m \right)$. Here we expressed the total energy of the ensemble in terms of its temperature: $E = s k_B T$ (and used $s \gg 1$), and assumed that the characteristic frequency $\omega_0$ is on the order of the bulk Debye frequency (indeed, the limiting cases – the vibrational frequency of the sodium dimer and $\Theta_D$ – differ only by a factor of 1.4 [7,65]).

Then the ratio of state densities can be rewritten as

$$\frac{\rho_N(E_N - D_N^m)}{\rho_N(E_N)} = e^{-D_N^m / k_B \tilde{T}_N^m} , \qquad (B.8)$$

where the logarithms of both terms on the left-hand side were expanded to first order in $D_N^m / 2$ about the midpoint $E_N - D_N^m / 2$ and [analogously to Eq. (B.5)] the effective temperature for monomer evaporation was introduced:

$$\frac{1}{k_B \tilde{T}_N^m} = \frac{d \ln \rho_N (E - \frac{1}{2} D_N^m)}{dE} \bigg|_{E = E_N} , \qquad (B.9)$$

If the cluster's heat capacity $C = dE/dT$ remains relatively constant in the energy interval from $E_N - D_N^m$ to $E_N$, the effective temperature can be expressed as

$$\tilde{T}_N^m\left(E_N\right) \approx T_N\left(E_N\right) - D_N^m / 2C .$$ (B.10)

Recall that $T_N(E_N) \equiv T_{parent}$. The fact that effective temperature $\tilde{T}_N^m$ used in the rate constant is lower than the temperature of the parent cluster is sometimes called the "finite heat bath correction."[66]

In the oscillator approximation for cluster ion vibrations, the heat capacity is

$$C \approx (3N-7)k_B.$$ (B.11)

A more familiar expression for a set of harmonic oscillators may be $C \approx (3N-6)k_B$, but the additional $k_B$ on the right-hand side derives from the fact that here we are dealing with the microcanonical, rather than canonical, heat capacity.[66]

For hot, especially molten, clusters this model is not very accurate. However, as shown in Refs 54,56 for alkali metals the inaccuracies appear to compensate each other and yield results close to those given by more quantitatively elaborate models.

Finally, combining these expressions with Eq. (B.7), we express the atom evaporation rate constant as follows:

$$k_N^m\left(E_N\right) = \Omega^m \left(\frac{\Theta_D}{T_{N-1}}\right) N^{2/3} e^{-D_N^m / k_B \tilde{T}_N^m} ,$$ (B.12)

where

$$\Omega^m = \frac{2Mr_s^2\left(k_B \Theta_D\right)^2}{\pi \hbar^3} .$$ (B.13)

The exponential is much more sensitive to temperature changes than the prefactor, so the latter is commonly viewed as a constant, $\omega^m = \Omega^m(\Theta_D/T_{N-1})$, and the expression is reduced to a pure Arrhenius form:

$$k_N^m(E_N) \approx \omega^m N^{2/3} e^{-D_N^m/k_B \tilde{T}_N^m} \; . \tag{B.14}$$

This has the same appearance as the rate constants deduced by the RRK method,[58,67] which is still sometimes encountered in the literature, but a key difference is that in the latter method the prefactor is assumed equal to the vibrational frequency, which leads to inaccuracies. For example, Fig. 8 illustrates the atomic evaporation rate constants for $Na_{23}$ calculated using expressions (B.12), (B.14), and the RRK assumption. It is seen that the use of a temperature-independent prefactor $\omega^m$ does not alter the rate constant significantly, but values calculated with $\omega^{RRK}$ are too low by a factor of $\sim 10^2$.

*Dimer evaporation*. The additional ingredient needed to obtain the dimer evaporation rate constant is the density of the internal degrees of freedom of the fragment. It is the product of the rotational and vibrational energy state densities of the dimer, so in Eq. (B.2) we write $\rho^d(\varepsilon_{\text{int}}^d) = \rho_r^d(\varepsilon_r^d)\rho_v^d(\varepsilon_v^d)$, $\varepsilon_{\text{int}}^d = \varepsilon_r^d + \varepsilon_v^d$. The sodium dimer ground state is a singlet so $g^d=1$; $\mu^d \approx 2M$, and the sticking cross section remains approximately the same.

The daughter state density factor $\rho_{N-2}\left(E_N - D_N^d - \varepsilon_r^d - \varepsilon_v^d - \varepsilon_t\right)$ can be expanded to first order in the dimer energies $\varepsilon_r^d, \varepsilon_v^d, \varepsilon_t$, similar to Eq. (B.4). Integration over these variables produces, in addition to the previously encountered integral over $\varepsilon_t$, also the integrals $\int \rho_r^d(\varepsilon_r^d)\exp\left(-\varepsilon_r^d/k_B T_{N-2}\right)d\varepsilon_r^d$ and $\int \rho_v^d(\varepsilon_v^d)\exp\left(-\varepsilon_v^d/k_B T_{N-2}\right)d\varepsilon_v^d$ which can be recognized as, respectively, the partition functions of vibrational and rotational states of the dimer fragment, $Z_v^d(T_{N-2})$ and $Z_r^d(T_{N-2})$. In the high-temperature limit these are equal to[68] $Z_v^d(T) = k_B T/(h\nu)$, $Z_r^d(T) = k_B T / (2B)$, where $\nu$ is the vibrational frequency and $B$ is the rotational constant of the dimer. Since, as mentioned above, the vibrational frequencies of the sodium dimer and the bulk are close, we can replace $h\nu$ by the Debye temperature: $Z_\nu^d(T_{N-2}) \approx T_{N-2} / \Theta_D$.

Proceeding now exactly as in the monomer case, the expression for the dimer evaporation rate constant becomes:

$$k_N^d(E_N) = \Omega^d \left( \frac{\Theta_D}{T_{N-2}} \right)^2 N^{2/3} e^{-D_N^d / k_B \tilde{T}_N^d} , \qquad (B.15)$$

where $\Omega^d = \dfrac{2M r_s^2 (k_B \Theta_D)^3}{2\pi \hbar^3 B}$ and

$$\tilde{T}_N^d(E_N) \approx T_N(E_N) - D_N^d / 2C . \qquad (B.16)$$

As in the monomer case, the prefactor in Eq. (B.15) varies relatively slowly and can be approximated by a constant, with the equation again taking the simple form

$$k_N^d(E_N) \approx \omega^d N^{2/3} e^{-D_N^d / k_B \tilde{T}_N^d} . \qquad (B.17)$$

It is essential to note, though, that because of the high density of dimer states the prefactor $\omega^d$ is much larger than $\omega^m$ for monomer evaporation. Indeed, using[65] $B$=0.15 cm$^{-1}$ and $T \sim 400$ K we see that

$$\frac{\omega^d}{\omega^m} \approx \frac{k_B \Theta_D}{2B} \frac{\Theta_D}{T} \approx 100. \qquad (B.18)$$

**Appendix C: Temperature distribution of clusters in the evaporative ensemble**

Clusters produced by the supersonic source are formed in the vicinity of the nozzle out of metal vapor which condenses within an expanding jet of carrier gas. After the transition to molecular flow, the clusters are initially hot enough to evaporate monomer and dimer fragments, and the basic tenet of the evaporative ensemble picture is that the abundance distribution downstream from the source arises from a series of post-nucleation evaporation steps. Each evaporation reduces the cluster's internal energy by an amount

$$\Delta E = D + 2k_B T_{daughter} + E_{int}, \qquad (C.1)$$

where the first term on the right-hand side is the dissociation energy, the second term is the translational kinetic energy of the evaporated fragment [Eq. (B.6)], and the last term is the internal energy of the fragment ($E_{int}$=0 for monomers and $E_{int}$=2$k_B T_{daughter}$ for dimers,[69] composed of two rotational degrees of freedom with ½$k_B T$ of energy each, and one vibrational degree of freedom with $k_B T$).

For cluster sizes considered here, the amount $\Delta T \approx \Delta E / C$ by which the particle temperature drops after evaporation, ranges from tens to hundreds of K. As a result, as was illustrated in Fig. 3(a), every subsequent evaporation step proceeds with a significantly lower rate constant (i.e., takes much longer). This fact is very convenient, as it allows us to approximately set the entire time of the evaporative sequence, i.e., the entire free flight time of a cluster in the beam, equal to the duration of its last completed evaporation act. This makes it straightforward to estimate $F_N^{(neutral)}$, the temperature distribution of precursor clusters in the beam.

***Monomer evaporation***. If only the monomer evaporation channel is open, the evaporation probability function [Eq. (3) with $k_N^d(T) = 0$, see the right-hand curve in Fig. 9(a)] exhibits a step-like shape with a pronounced threshold temperature. This threshold defines the maximum temperature of clusters of size $N$ in the beam. Indeed, if a cluster has a higher temperature it will proceed to evaporate a fragment, whereas below this temperature it can survive long enough to be detected. We define the threshold as the inflection point of the rise: $P_N^m(T_N^{\max}, t_{flight}) = 1/e$. Here $t_{flight}$ is the beam flight time, as discussed in the previous paragraph; in the present case $t_{flight}$ is the travel time of the neutral clusters from the nozzle to the electron gun, measured to be $\approx 2$ ms. Then, employing Eq. (B.14) in the exponent of Eq. (3), we arrive at the following expression for the maximum cluster temperature in the evaporative ensemble:

$$T_N^{\max} \approx \frac{D_N^m}{k_B \ln\left(\omega^m t_{flight} N^{2/3}\right)} + \frac{D_N^m}{2C}. \tag{C.2}$$

The lower temperature limit, in turn, arises from the fact that every cluster is itself a product of an in-flight evaporation process. Therefore its parent's temperature had to exceed its own evaporation threshold. Relating the temperature of the parent to that of the daughter, we can write

$$T_N^{\min} \approx T_{N+1}^{\max} - \frac{\Delta E}{C}. \tag{C.3}$$

The parent's decay probability is also a sharp function of its temperature, translating into a step-like lower boundary of the daughter's temperature distribution [see the left-hand curve in Fig. 9(a)]. It is therefore a sensible approximation to treat $F_N^{(neutral)}(T)$ as a uniform distribution confined between $T_N^{\min}$ and $T_N^{\max}$, as stated in Sec. V.B and illustrated in Fig. 9(a).

If we neglect the difference between $D_{N+1}^m$ and $D_N^m$, the small shifts in $\ln(N^{2/3})$ and in the heat capacity, and the fragment kinetic energy in Eq. (C.1), then we see that the cluster temperature range is approximately

$$\overline{T}_N \pm \Delta T_N \approx \frac{D_N^m}{k_B[G + \frac{2}{3}\ln N]} \pm \frac{D_N^m}{2C} . \tag{C.4}$$

The quantity $G = \ln(\omega^m t_{flight})$ is known as the "Gspann parameter," and its typical magnitude in cluster beam experiments is $G\sim25\text{-}30$.[70] The physical meaning of the expression for $\overline{T}_N$ lies in the argument, originally formulated by Gspann,[71] that the temperature of clusters detected in a beam apparatus should be such, that the hottest have an evaporation lifetime $[1/k_N^m$, Eq. (B.14)] of about the flight time $t_{flight}$.

This expression includes only the lowest-order variation of the dissociation energy. For evaluating metastable decay fractions[72] or abundance variations,[17] it is essential to include higher-order terms into the approximate expression for $\Delta T_N$ in Eq. (C.4).

***Dimer evaporation***. The picture gets considerably more complicated when a cluster can originate from not one, but two precursors, either by monomer or dimer evaporation. For sodium clusters in the size range under study this occurs only for those with an even number of electrons: according to the data in Ref. 9 only even-numbered clusters have a substantial probability of evaporating a dimer.

The upper boundary of a cluster's temperature distribution is still defined by Eq. (C.2) (employing the most probable evaporation pathway of the cluster), but each of its parents now provides a separate lower temperature value. That is to say, the temperature distribution is a superposition of two distinct distributions with two different lower limits. It is difficult to precisely evaluate the relative weights of these distributions, because they depend on the relative abundances of the $N+1$ and $N+2$ parents, as well as on the ratio of their evaporation probabilities which itself is temperature-dependent (cf. Fig. 4). Hence one has to construct a usable qualitative approximation.

Based on the picture that many clusters undergo two successive rapid evaporations, we assume that the number of $N+2 \rightarrow N+1$ and the number of $N+1 \rightarrow N$ transitions are of the same order of magnitude, and take them to be roughly equal. Then the weights of the two superimposed temperature distributions for the cluster of size $N$ become just the branching ratios of the two evaporation channels of the $N+2$ parent. This is illustrated in Fig. 9(b). Clearly, this is only a schematic prescription, but it does have the advantage of, on one hand reflecting the general character of the evaporation sequence, and on the other hand being tractable. As mentioned in the text, the branching ratios were taken from the data in Ref. 9.

## Appendix D:  Dissociation energies of cation clusters

As described in the text and in Appendix E below, the dissociation energies of neutral and anionic clusters employed in our calculations were derived from the well-known experimental data set of cationic dissociation energies by Bréchignac *et al.,* Ref. 9.   In their experiment, internally excited sodium cations were produced by multiphoton ionization, and allowed to undergo a free-flight decay for a fixed period of time.   By measuring the intensities of the resulting evaporation products, the authors deduced the monomer and dimer evaporation rate constants and branching ratios, and finally extracted both types of dissociation energies for the ions by using the expression[73]

$$k_N^{(+)m\{d\}}(E_N) = 8\pi\mu g\sigma v_0^3 (3N-7) \frac{\left(E_N - D_N^{(+)m\{d\}}\right)^{3N-8}}{E_N^{3N-7}}, \tag{D.1}$$

where $2\pi v_0 = \omega_0$ is the vibrational frequency of ions and all other quantities have the same meaning as in Appendix B.   It is easy to see that this expression follows from Eq. (B.7) if one uses the Kassel state density of an ensemble of identical harmonic oscillators (see Appendix B). For $3N$-8>>1 and $D<E$, expanding the numerator recovers the Arrhenius form (B.14).

As discussed in Appendix B, this formula is adequate for the monomer evaporation rate constants, but it neglects to properly account for the rotational and vibrational states of the dimer. As a result, the correct pre-exponential factor for dimer cation evaporations differs from that used for the determination of $D_N^{(+)d}$ in Ref. 9 by a factor of ~$10^2$.  This introduces a systematic inaccuracy, on the order of 20%,[33] into the resulting dissociation energy values.

Unfortunately, the inaccuracy cannot be fixed by simply replacing the incorrect prefactor ($\omega^m$) by the correct one ($\omega^d$), and shifting $D_N^{(+)d}$ in the exponential in Eq. (B.17) to compensate.

The reason is that in Ref. 9 the dissociation energies for both monomer and dimer loss were fitted simultaneously to the cation dissociation profiles while constrained to fulfill the Born-Haber cycle condition $D_N^{(+)m} - D_N^{(+)d} + D_{N-1}^{(+)m} = D_2$, where $D_2$ is the dimer formation energy. This means that the monomer and dimer dissociation energy fits are intertwined and have to be readjusted simultaneously.

Since it is unfeasible to reevaluate the dissociation energies from the raw experimental spectra in Ref. 9, the only practical and consistent solution for the present analysis was to employ the tabulated energy values $D_N^{(+)m}$ and $D_N^{(+)d}$ together with the prefactor $\omega^m$ for both.[74] Because these quantities were matched in the original work so as to reproduce its evaporation kinetics, it did not come as a surprise that they provided a satisfactory description of anion evaporation cascades as well. For use in the calculations, the cation dissociation energies were converted into the corresponding parameters for neutral and anionic clusters by means of the correction formulas described in Appendix E.

## Appendix E:  Dissociation energies of neutral and anion clusters

Nanocluster dissociation energies $D_N$ are central quantities in a statistical evaporation analysis:  they define the lifetimes and the temperature distributions of the fragmenting particles. For the purposes of our experiment, we need to know the monomer and dimer dissociation energies of sodium neutral and anion clusters, but experimental data exist only for cations (see Appendix D).  Theoretical computations for anions are available only for the doubly-charged $Na_N^{(2-)}$.[76]

A common first approximation for monomer evaporation is to write $D_N^m \approx D_{N+1}^{(+)m} \approx D_{N-1}^{(-)m}$ (see, e.g., Ref. 44).  In the following, we refine this approximation by using the ideology of the shell-correction method (see, e.g., the reviews in Refs. 77-80).  The total binding energy of a metal cluster with $N$ ion cores and $n$ valence electrons is expressed as a sum:

$$E_{tot}(N,n) = \bar{E}(N) + \delta E(n). \tag{E.1}$$

The first term is a smooth function of the cluster radius.  For neutral clusters it is defined in terms of a liquid drop expansion:

$$\bar{E}^{(neutral)}(N) \approx a_v N - a_s N^{2/3}, \tag{E.2}$$

where $a_v$=1.12 eV and $a_s$=1.02 eV are the volume and surface energy coefficients for sodium clusters.[10].  The second term in Eq. (E.1) is the oscillating quantum shell correction.  Two assumptions are made: (1) that $\delta E$ is determined only by the degree of shell filling and hence by the total valence electron count $n$, and is independent of the cluster charge, and (2) that the droplet energy difference between two particles of the same size but different charge states is determined by the electrostatics of a spherical droplet.  Specifically, one writes

$$\bar{E}^{(+)}(N) = \bar{E}^{(neutral)}(N) + I_N = \bar{E}^{(neutral)}(N) + \left( W + \frac{3}{8}\frac{e^2}{R_N} \right) \tag{E.3}$$

and

$$\overline{E}^{(-)}(N) = \overline{E}^{(neutral)}(N) - A_N = \overline{E}^{(neutral)}(N) - \left( W - \frac{5}{8} \frac{e^2}{R_N} \right),$$ (E.4)

where $I$ is the ionization potential, $A$ is the electron affinity, and $W$ is the bulk work function.

The semiclassical expressions in parentheses for $I$ and $A$ have been extensively discussed in the literature, see, e.g., Refs. 10,81. The experimental values of these parameters for simple metal clusters (see the compilations and analyses in Refs. 82,83), show a good general agreement with such scaling. Slight deviations of the asymptotic values from $W$ are not important for the present purposes, as this constant cancels out in the calculations below, while deviations of the coefficients from the values 3/8 and 5/8 do not modify the formulas derived below to a significant degree.[84]

The monomer dissociation energy for a neutral cluster can be derived through the following manipulations:

$$\begin{aligned}
D_N^m &= E_{tot}^{(neutral)}(N) - E_{tot}^{(neutral)}(N-1) \\
&= \overline{E}^{(neutral)}(N) - \overline{E}^{(neutral)}(N-1) + \delta E(N) - \delta E(N-1) \\
&= \overline{E}^{(neutral)}(N) - \overline{E}^{(neutral)}(N-1) + \left[ E_{tot}^{(+)}(N+1) - \overline{E}^{(+)}(N+1) \right] - \left[ E_{tot}^{(+)}(N) - \overline{E}^{(+)}(N) \right] \\
&= D_{N+1}^{(+)m} + \overline{E}^{(neutral)}(N) - \overline{E}^{(neutral)}(N-1) - \left[ \overline{E}^{(neutral)}(N+1) + I_{N+1} \right] + \left[ \overline{E}^{(neutral)}(N) + I_N \right] \\
&= D_{N+1}^{(+)m} - \Delta_2 \left[ \overline{E}^{(neutral)}(N) \right] + \left( I_N - I_{N+1} \right),
\end{aligned}$$

(E.5)

where $\Delta_2$ is the second difference of the liquid drop binding energy, here equal to the second derivative of Eq. (E.2).

Expanding the differences in powers of $1/N$ we obtain[85]

$$D_N^m = D_{N+1}^{(+)m} + \frac{e^2}{8NR_N} - \frac{2a_s}{9N^{4/3}},$$ (E.6)

and a similar calculation for anions yields

$$D_N^{(-)m} = D_{N+1}^m + \frac{5e^2}{24NR_N} - \frac{2a_s}{9N^{4/3}} \ . \tag{E.7}$$

These expressions supply the leading-order ($N^{4/3}$) smooth corrections to the relation between the monomer dissociation energies in neutral and ionic clusters.

For dimer evaporation the treatment is the same, except for cluster size differentials equal to $2N$ instead of $N$. As a result, the corrections increase by a factor of two:

$$D_N^d = D_{N+1}^{(+)d} + \frac{e^2}{4NR_N} - \frac{4a_s}{9N^{4/3}} \ , \tag{E.8}$$

$$D_N^{(-)d} = D_{N+1}^d + \frac{5e^2}{12NR_N} - \frac{4a_s}{9N^{4/3}} \ . \tag{E.9}$$

The sodium cluster dissociation energies employed in our evaporative ensemble calculations were derived from the original measurements of Ref. 9 (see Appendix D) using Eqs. (E.6)-(E.9).

**Figure Captions**

**Fig. 1.** Outline of the experiment. A beam of neutral clusters was created by supersonic expansion through a 75 μm nozzle. The source body was kept at 660°C and the Ar carrier gas pressure varied from 300–600 kPa. The electron gun intersected the cluster beam (collimated to 1.4 mm×1.4 mm) by a ribbon of slow electrons (1.4 mm×25.4 mm, ~10 μA; a magnetic field of 400 Gauss collinear with the electron current was used to counteract its dispersal by space-charge effects). The negatively charged products were extracted by an electrostatic lens, filtered by a quadrupole mass analyzer (QMA), and detected by a channeltron. The remaining neutral clusters were ionized by a uv lamp and detected by another QMA. In this way, the abundance spectra of the precursor and anion clusters were recorded simultaneously.

**Fig. 2**. Mass spectra of (a) the electron attachment products and (b) of the precursor beam. The different shades in (a) represent separate segments for which the cluster source was optimized for maximum precursor intensity and the first QMA was adjusted for the strongest mass-resolved anion signal. Note that the anion pattern displays not only a magic number shift relative to the neutral precursors, but also a strong alteration in the relative abundances of open-shell clusters.

**Fig. 3.**  (a) A diagram of the evaporative cooling process.  As the hot cluster evaporates atoms (as in this sketch) or dimers, its internal energy decreases and the interval before the next evaporation step increases exponentially.  The "final" temperature corresponds to the point when the lifetime becomes much greater than the beam flight time.

(b) An outline of the anion fragmentation chain constructed out of monomer and dimer evaporation steps.  Those "daughters" which have sufficiently low internal temperatures to survive on the detection time scale are circled.  The end products of all such evaporation cascades add up to form the measured anion abundance distribution in Fig. 2(a).

**Fig. 4.**  Dashed lines: monomer and dimer evaporation probabilities for a model cluster, assuming $N$=16 atoms, $D^m$=0.8 eV, $D^d$=0.9 eV and a flight time of 2 ms, calculated using Eq. (3) and Appendix B as a function of the cluster's internal temperature.  Solid line: the probability that the parent cluster does *not* evaporate during the flight time, given by $\exp\left[-k^m(T) - k^d(T)\right]$.

**Fig. 5**.  Internal temperature distributions of sodium clusters immediately following electron attachment.  The boundaries of the distributions are obtained from those of the neutral $Na_N$ precursors, modeled corresponding to the discussion in Appendix C, by shifting them upwards according to Eq. (5).

**Fig. 6**.    Relative abundance distributions of $Na_N^-$ clusters formed by low-energy electron attachment.    The two panels correspond to data segments acquired under different optimization conditions, as marked in Fig. 2.    The energy released by the captured electron is thermalized, and subsequent cluster cooling via evaporation of atoms and dimers restructures the abundance spectrum.    The modeled distribution was derived by convoluting the mass spectrum of the precursor neutral beam first with the electron attachment cross sections and then with the systematically evaluated evaporation pathway probabilities.    For plotting, the relative intensity distributions were scaled so as to achieve overlap (note that for $Na_{19}$ different scaling was used in the two segments).

**Fig. 7.**  Insert: cation mass spectrum obtained by ionizing the beam of neutral sodium clusters with coaxial 350 nm laser pulses.  Open bars:  cluster mass spectrum measured by above-threshold UV lamp ionization, as in Fig. 2(b).  Solid bars: neutral cluster abundance distribution derived, as described in the text, by adjusting the UV-lamp data with the aid of the laser-ionization data.

**Fig. 8.**  Temperature dependence of the rate constant for monomer (i.e., atom) evaporation by the $Na_{23}$ cluster calculated with the use of Eq. (B.12) [dashed line; $\Omega^m=4\times10^{15}$ $s^{-1}$, $\Theta_D=150$ K, $D_{23}^m=0.74$ eV] or (B.14) [solid line; $\omega^m\approx\Omega^m(\Theta_D/500$ K$)=1.2\times10^{15}$ $s^{-1}$].    The RRK

assumption that the Arrhenius prefactor is of the order of the vibrational frequency [dotted line; $\omega^{\text{RRK}} \approx 3 \times 10^{13}$] yields significantly lower rate constants.

**Fig. 9.** (a) In the evaporative ensemble picture, the boundaries of cluster evaporation probabilities (solid lines) define the region of observable cluster temperatures, as discussed in Appendix C. The cluster temperature distribution function $F_N(T)$ will be approximated by a (normalized) rectangular-box distribution with boundaries denoted by the dashed lines. The calculation shown is for $Na_{17}$, representing clusters which can originate only from a single parent by monomer evaporation (here, $Na_{18}$). (b) Clusters which can be produced by both: monomer and dimer evaporation channels are assigned a double-step internal temperature distribution. Analogously to the diagram in part (a), the upper boundary $T^{max}$ is defined by the most probable evaporation pathway of the cluster, and the lower boundaries correspond to the decay probabilities of its two parents. The weights are approximated by the monomer and dimer evaporation branching ratios of the larger parent, $W_{N+2}^{m}$ and $W_{N+2}^{d}$.

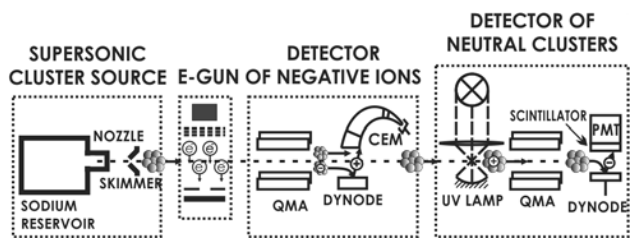

**FIG. 1**

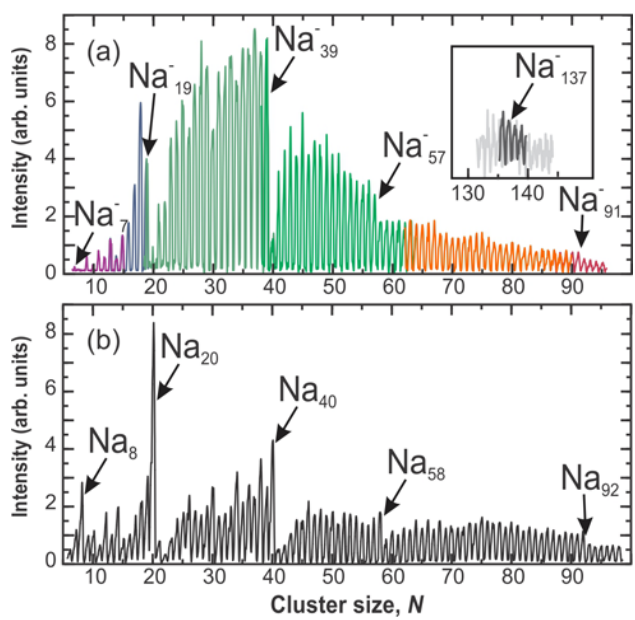

**FIG. 2**

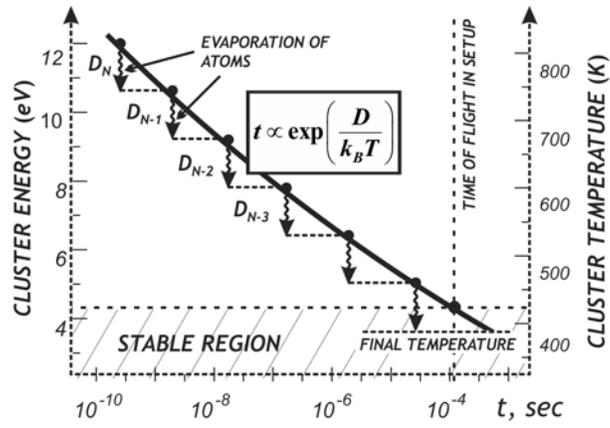

(a)

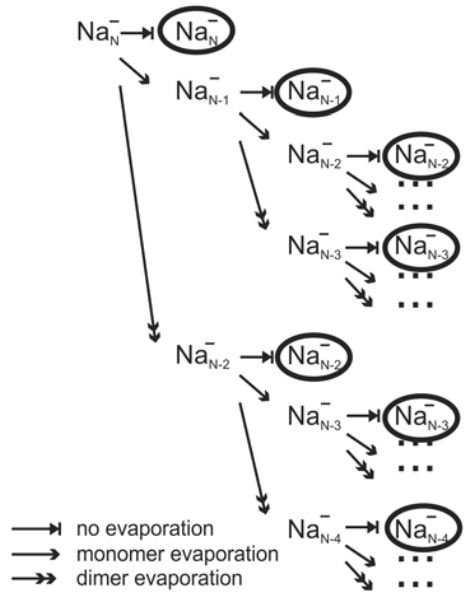

- no evaporation
- monomer evaporation
- dimer evaporation

(b)

**FIG. 3**

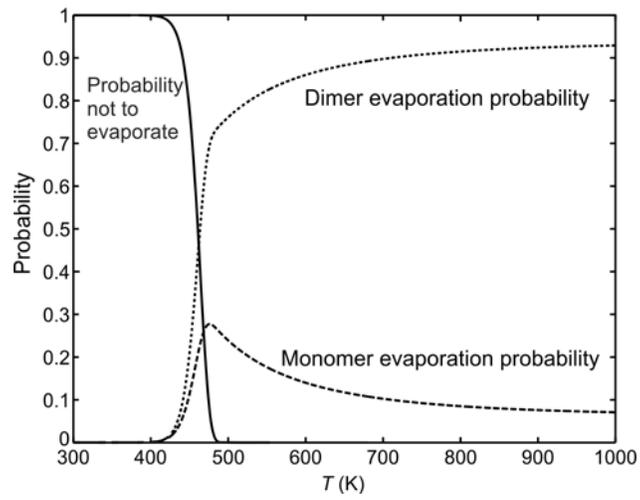

**FIG. 4**

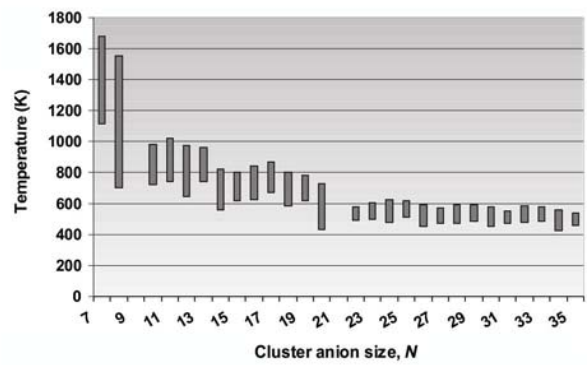

**FIG. 5**

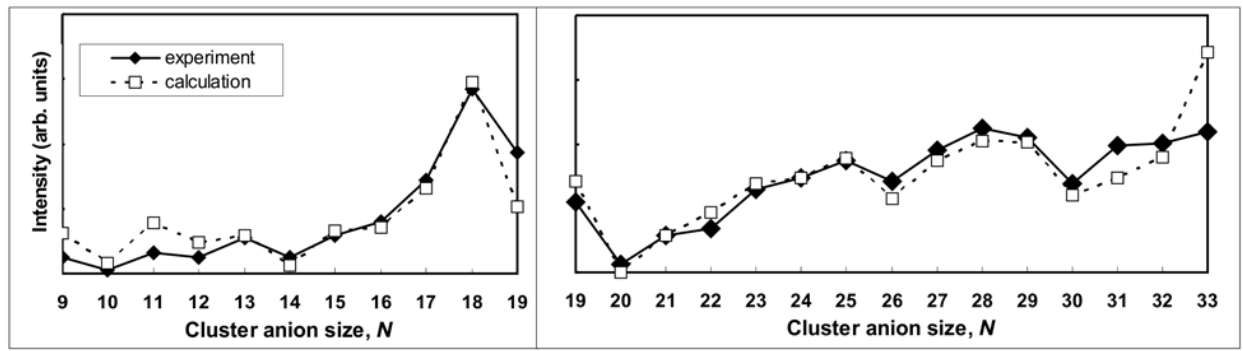

**FIG. 6**

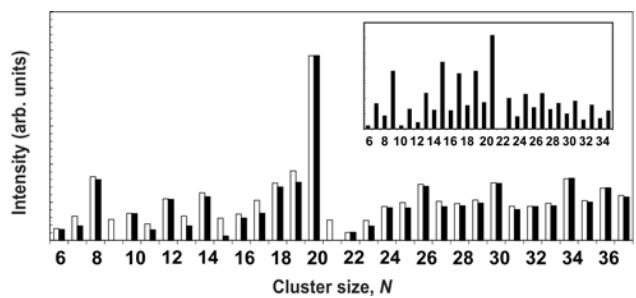

**FIG. 7**

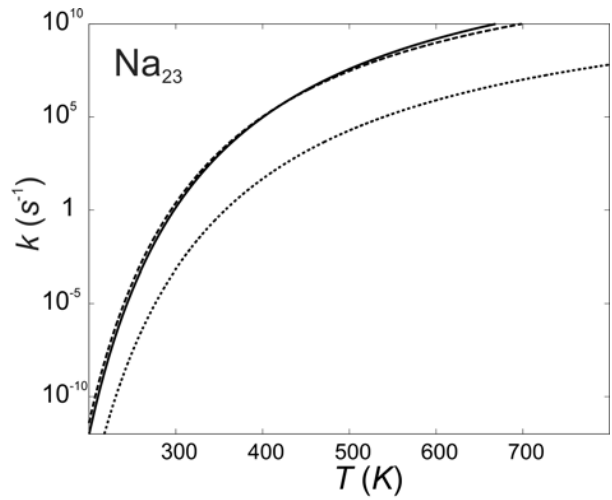

**FIG. 8**

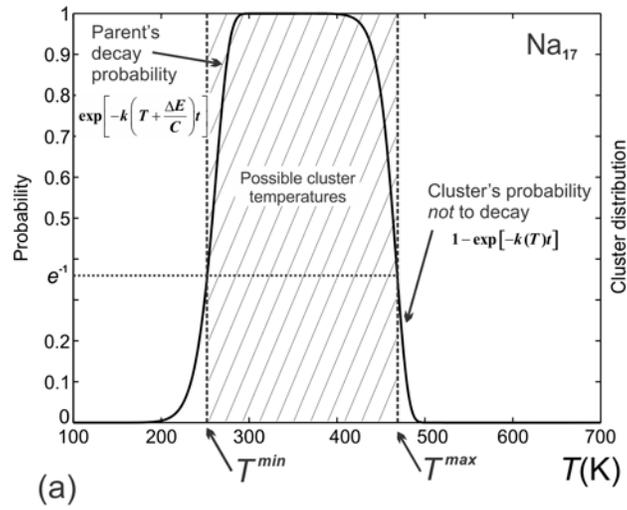

(a)

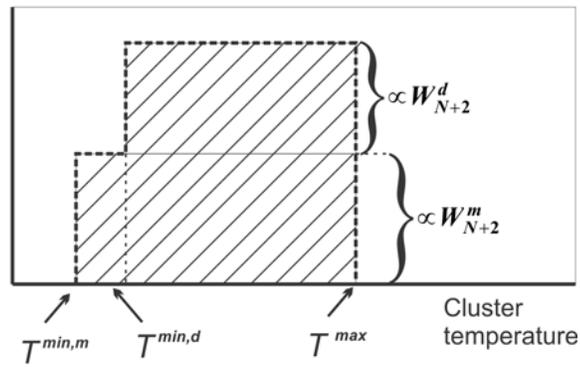

(b)

**FIG. 9**